\documentclass[a4paper]{article}
\usepackage[latin1]{inputenc}
\usepackage{graphicx}
\usepackage{amssymb}
\usepackage{amsfonts}
\usepackage[all]{xy}
\usepackage[thmmarks]{ntheorem}

\newcommand{\eqref}[1]{(\ref{#1})}

\def\A{{\mathcal A}}

\def\K{{\mathcal K}}
\def\D{{\mathcal D}}

\def\B{{\mathcal B}}

\def\RR{{\mathbb R}}
\def\CC{{\mathbb C}}

\def\HH{{\mathbb H}}

\def\End{\mbox{\rm End}}

\def\Aut{\mbox{\rm Aut}}

\def\bra{\langle}
\def\ket{\rangle}
\def\id{{\rm Id}}

\def\diag{{\rm diag}}

\def\bea{\begin{eqnarray}}
\def\eea{\end{eqnarray}}

\def\Spin{{\rm Spin}}

\def\bea{\begin{eqnarray}}
\def\eea{\end{eqnarray}}

\def\be{\begin{equation}}
\def\ee{\end{equation}}

\newenvironment{rem}[1][{}]{\smallbreak \noindent  {\bf Remark #1}\small }

\newtheorem{theorem}{Theorem}
\newtheorem{definition}{Definition}
\newtheorem{lemma}{Lemma}
 
\newtheorem{propo}{Proposition}
\theoremstyle{nonumberplain}
\theorembodyfont{\normalfont}
\theoremseparator{:}
\theoremsymbol{$\P$}
\newtheorem{demo}{Proof}
\begin{document}
\title{On the uniqueness of Barrett's solution to the fermion doubling problem in Noncommutative Geometry}
\author{Fabien Besnard\footnote{P\^ole de recherche M.L. Paris, EPF, 3 bis rue Lakanal, F-92330 Sceaux.}\\ {\small fabien.besnard@epf.fr}}
\maketitle
\begin{abstract}
A solution of the so-called fermion doubling problem in Connes' Noncommutative Standard Model has been given by Barrett in 2006 in the form of Majorana-Weyl conditions on the fermionic field. These conditions define a ${\cal U}_{J,\chi}$-invariant subspace of  the correct physical dimension, where ${\cal U}_{J,\chi}$ is the group of Krein unitaries commuting with the chirality and real structure. They require  the KO-dimension of the total triple to be $0$. In this paper we show that this solution is the only one with this invariance property, up to some trivial modifications, and under some very natural   assumptions on the finite triple.  We also observe that a simple modification of the fermionic action can act as a substitute for the explicit projection on the physical subspace.
\end{abstract}

\section{Introduction}
What are the advantages of the Noncommutative Geometry approach to particle physics with respect to the traditional approach ? To answer this question in the most concise way, there are two:
\begin{enumerate}
\item Some features which are added by hand in the usual formulation of the Standard Model (SM) pop up naturally in the Noncommutative Standard Model (NCSM): these include the Higgs field and the neutrino mixing term  \cite{SMmix}.
\item\label{point2} The NCSM is much more constrained.
\end{enumerate}
The constraints alluded to in \ref{point2}  come in particular\footnote{But not only: there are also constraints on the Dirac operator about which we will not talk in this paper. See for instance \cite{whySM}, \cite{BBB}.} from the fact that in the SM one starts with a Lie group, while in the NCSM  one starts with a finite-dimensional $*$-algebra. By the Artin-Wedderbun theorem, such an algebra is a direct sum of matrix algebras over the reals, complex numbers or quaternions. There is a much wider selection of simple Lie groups: already at the level of the Lie algebra there are four infinite families and five exceptional cases, and one must also take into account the topology of the group. Moreover, and perhaps more importantly, the representation theory of $*$-algebras is much simpler than that of groups: there is only one irreducible representation of $M_n(K)$, $K=\RR,\CC,\HH$, namely the regular one. 
However, the NCSM also has problems, in particular:

\begin{enumerate}
\item the unimodularity condition,
\item the definition of the spectral action in the Lorentzian setting,
\item the quadrupling of the fermionic degrees of freedom, hereafter called ``fermion doubling'' for historical reasons\footnote{It was called this way in \cite{LMMS} because it can be seen as the combination of two doublings of the degrees of freedom, among which only one, giving rise to mirror fermion-like terms in the action, were found to be worrisome.}.
\end{enumerate}

The first problem is maybe the most pressing one, but we won't deal with it in this paper. The second problem can be by-passed: the spectral action can be replaced by the older Connes-Lott action in the Lorentzian setting without trouble \cite{thesenadir}. The third problem is the main subject of this work. It has been first stressed in  \cite{LMMS}. Its technical origin  is the following. The spectral triple of the NCSM is a so-called ``almost-commutative manifold'', i.e. it is the tensor product of the spectral triple of the spacetime manifold and a finite spectral triple.  The latter includes a finite-dimensional space ${\cal K}_F$, a basis of which is $(p_\sigma)$ where $p$ runs over all  elementary particles and $\sigma$ is a symbol among $R,L,\bar R,\bar L$. For instance $e_{\bar L}$ is interpreted as an anti-left electron. However, the particle/antiparticle type and parity are already present in the spinor space $S_x$ at a point of the manifold. The space $S_x\otimes {\cal K}_F$ pertaining to the almost-commutative spectral triple of the NCSM thus has four times the expected dimension. It should be noted that the presence of the symbol $\sigma$ in the first place is necessary in order to have the correct representation of the gauge group on fermions. To solve this problem, one looks for a subspace $H_x$ of $S_x\otimes {\cal K}_F$ of the correct dimension, calls it the ``physical subspace'' and declares fermion fields to have values in this subspace (a kind of superselection rule). Clearly, the physical subspace must also be invariant under the gauge group.

In 2006, Barrett \cite{Barrett} noticed that a physical subspace could be defined by the conditions
\begin{equation}
J\Psi=\Psi,\qquad \chi\Psi=\Psi \label{BC}
\end{equation}
where $J$ is the real structure and $\chi$ the chirality of the total triple, henceforth called Majorana-Weyl or Barret's conditions. Since each one of the conditions \eqref{BC} divides the dimension by a factor of two, we see that they indeed solve  fermion ``doubling''. Note that fermion doubling seems to be much more than just a technical problem to solve, having connections with apparently unconnected matters, such as Wick rotations \cite{AKL} or neutrino mixing. To understand the latter point, which was observed in \cite{Barrett} and independently in \cite{SMmix} (thanks to an alternative technique available in Euclidean signature), note that equations \eqref{BC} are consistent only if $J^2=1$ and $J \chi=\chi J$. They thus require the total KO-dimension of triple to be $0$.  In $1+3$ spacetime dimension, this is equivalent to $J_F^2=-1$, where $J_F$ is the real structure of the finite triple.  By what seems to be a happy coincidence, precisely this sign allows for  a neutrino mixing term in the action (i.e. the neutrino mixing term must vanish if $J_F^2=1$). Given the reach of these discoveries, it seems important to inquire on the uniqueness of Barrett's solution in KO-dimension $0$, as well as on the existence of any solution at all in other KO-dimensions. To our knowledge these questions have not yet been considered in the literature.

Some observations are immediate. First,  one can introduce a phase and a sign, and require 
\begin{equation}\label{BCmodif}
J\Psi=e^{i\varphi}\Psi,\qquad \chi\Psi=\pm \Psi 
\end{equation}
instead of \eqref{BC}.  Moreover, the space defined by \eqref{BCmodif} is not only invariant by the gauge group of the Standard Model but by the larger group ${\cal U}_{J,\chi}$ of Krein unitaries commuting with $\chi$ and $J$. In this paper we will show that, with the assumption of invariance under ${\cal U}_{J,\chi}$, there exists no other solution than \eqref{BCmodif}, and in particular the KO-dimension has to be $0$. We will also remark that uniqueness is lost if one only assumes gauge invariance, though we will not fully investigate this much more complex case.   Barrett's solution thus seems to be a satisfactory and almost unique solution to the fermion doubling problem, assuming a very natural invariance. Nevertheless, conditions such as \eqref{BC} may appear as a feature added by hand to the beautiful construct of Noncommutative Geometry. However, we will observe that instead of projecting the fields on the physical subspace, the resolution of the fermion doubling problem can be achieved by changing the usual  fermionic action $S(A,\Psi)=(\Psi,D_A\Psi)$ to 
%
\begin{eqnarray}
S'(A,\Psi)=&\frac{1}{8}[(\Psi,D_A\Psi)+(D_A\Psi,\Psi)+(D_A\Psi,J\Psi)+(J\Psi,D_A\Psi)\cr
&+(\chi\Psi,D_A\Psi)+(D_A\Psi,\chi\Psi)+(D_A\Psi,\chi J\Psi)+(\chi J\Psi,D_A\Psi)]
\end{eqnarray}
%
Using $S'$ is equivalent to projecting fermionic fields on the physical subspace and then use the traditional $S$. The action $S'$ may seem even more natural than $S$, since it involves all the possible combinations of the background objects with which we can build the action, with only the simplest coefficients. With this point of view, the fermionic fields orthogonal to the physical subspace still exist, but do not participate in any interaction.

In this paper we will assume that the spacetime manifold is four-dimensional with metric signature $(1,3)$. In this setting a Connes-Lott type noncommutative gauge theory reproducing exactly the bosonic and fermionic actions of the SM can be constructed thanks to a certain finite \emph{indefinite} triple \cite{thesenadir}. This triple belongs to a family described in section 2 of this paper, which has $4$ free parameters: the signs $\epsilon_F$ and $\kappa_F$, and the components $\eta_R,\eta_L$ of the internal metric. This family is characterized by two conditions: the internal metric has a tensor product form, and the real structure exchanges particles and anti-particles. These are thus two additional (though very natural) hypotheses that we make. We will show in section 3 that a ${\cal U}_{J,\chi}-$invariant physical subspace exists  iff $\epsilon_F=-1$, and that in this case it is given by  \eqref{BCmodif}. The  parameter $\kappa_F$  is  then determined   by the requirement to recover a non-trivial fermionic action, as we recall in section 4.  In section 5, we will conclude the paper by observing that in the case singled out by the previous considerations, the projection on the physical subspace can be replaced with the use of the action $S'$.

We stress that in order to write down precise mathematical statements, we are forced to set ourselves in a specific framework with respect to the definition of indefinite spectral triples, while it is a still evolving subject. We will use the one proposed in \cite{algback}, which will be reviewed in section 2. However, our results are largely independent from the details of the formulation.

\section{General setting}

We will use the general setting of algebraic backgrounds, proposed in  \cite{algback}. However, no familiarity with this notion is required, since we recall everything we need below. 

A  \emph{pre-Krein space} is a complex vector space $\K$ equipped with a non-degenerate indefinite metric $(.,.)$, which is decomposable into the direct sum $\K=\K_-\oplus \K_+$ of a positive and negative definite subspaces. Giving any such decomposition is equivalent to giving a fundamental symmetry $\eta$, which satisfies $\eta^\times=\eta$ (where $\times$ is the adjoint with respect to $(.,.)$), $\eta^2=1$, and such that $\bra .,.\ket_\eta:=(.,\eta.)$ is positive definite. It is said to be ${\mathbb Z_2}$-graded and real if there exists a linear operator $\chi$ (chirality) and an antilinear operator $J$ (graded real structure) such that
\be 
\chi^2=1,\quad J^2=  \epsilon,\quad J\chi=  \epsilon''\chi J,\quad J^\times  =  \kappa J,\quad \chi^\times =  \epsilon''  \kappa''\chi\label{kosigns}
\ee
where $\epsilon,\kappa,\epsilon'',\kappa''$ are signs (``KO-metric signs''). A fundamental symmetry $\eta$ is said to be compatible with $\chi$ and $C$ iff 
\be
\chi\eta=\epsilon''\kappa''\eta\chi\mbox{ and }J\eta=\epsilon\kappa \eta J\label{compsigns}
\ee

The pre-Krein space $\K$ can be decomposed into even and odd subspaces, $\K=\K_0\oplus \K_1$, which are the eigenspaces of $\chi$. An operator $A$ which commutes with $\chi$ will respect this decomposition and will be called \emph{even}. If $A$  anticommutes with $\chi$ it will exchange $\K_0$ and $\K_1$ and be called \emph{odd}. Note that if $\epsilon''\kappa''=1$ then $\chi^\times=\chi$ and this implies that $\K_0$ and $\K_1$ are orthogonal with respect to $(.,.)$. In this case we will say that the Krein product is even. On the contrary if $\epsilon''\kappa''=-1$, $\K_0$ and $\K_1$ are self-orthogonal ($\K_i=\K_i^\perp$) and we say that the Krein product is odd. These considerations will be of particular importance when we turn to tensor products.

We recall that  $\epsilon,\epsilon''$ are given in terms of $n$,  an integer modulo $8$ called the KO-dimension, by the formulas $\epsilon=(-1)^{n(n+2)\over 8}$, $\epsilon''=(-1)^{n/2}$, while $\kappa=(-1)^{m(m+2)\over 8}$, $\kappa''=(-1)^{m/2}$, where $m$ is another integer modulo $8$ called the metric dimension (for more details see \cite{BBB}).
For convenience the values of the signs $\epsilon,\epsilon'',\kappa,\kappa''$ in terms of $m,n$ are gathered in table \ref{tab1}.
\begin{table}[hbtp]
\begin{center}
\begin{tabular}{|c||c|c|c|c|}
\hline
m,n&0&2&4&6\\
\hline
$\kappa,\epsilon$&1&-1&-1&1\\
\hline
$\kappa'',\epsilon''$&1&-1&1&-1\\
\hline
\end{tabular}
\end{center}
\caption{ Signs $\epsilon,\epsilon'',\kappa,\kappa''$ in terms of $m,n$.}\label{tab1}
\end{table}
\begin{definition}\label{algbgd} An \emph{algebraic background}   is a tuple ${\cal B}=(\A, \K, (.,.),\pi,\chi,J,\Omega^1)$  where:
\begin{enumerate}
\item$(\K,(.,.),\chi,J)$ is a ${\mathbb Z}_2$-graded real pre-Krein space, 
\item $\A$ is a $*$-algebra and $\pi$ is  a  faithful  $*$-representation of it by even operators on $\K$,
\item the ``bimodule of 1-forms'' $\Omega^1$ is an  ${\cal A}$-bimodule of odd operators on $\K$.
\end{enumerate}
\end{definition}
\begin{rem} In the general case, $\A$ is not required to be a $*$-algebra, as explained in \cite{algback}. However, since it will  always be the case in this paper, we prefer to put it in the axioms. There are also some boundedness conditions on $\pi(\A)$ and $\Omega^1$, but we do not need to enter into these details.
\end{rem}

Given $\B$, one can define its \emph{configuration space} $\D_\B:=\{D\in\End(\K)|D^\times=D, \chi D=-D\chi, JD=DJ,$ and $\forall a\in \A,\ [D,\pi(a)]\in\Omega^1\}$. The elements of $\D_B$ are called the \emph{compatible Dirac operators}. The group of automorphisms of $\B$ is $\Aut(\B):=\{U\in\End(K)|UU^\times=1, UJ=JU, U\chi=\chi U, U\Omega^1U^{-1}=\Omega^1, U\pi(\A)U^{-1}=\pi(\A)\}$. Of crucial importance for us is the definition of the tensor product.

\begin{definition}
Let ${\cal B}_1=({\cal A}_1,{\cal K}_1,D_1,J_1,\chi_1,\Omega^1_1)$ and ${\cal B}_2=({\cal A}_2,{\cal K}_2,D_2,J_2,\chi_2,\Omega^1_2)$ be two algebraic backgrounds. Then the tensor product algebraic background ${\cal B}={\cal B}_1\hat\otimes {\cal B}_2$ is defined by:
\begin{itemize}
\item ${\cal A}={\cal A}_1\otimes {\cal A}_2$ with involution $(a_1\otimes a_2)^*=a_1^*\otimes a_2^*$ and representation $\pi(a_1\otimes a_2)=\pi_1(a_1)\otimes\pi_2(a_2)$.
\item ${\cal K}={\cal K}_1\hat \otimes {\cal K}_2$   tensor product with indefinite product
\begin{equation}
(\phi_1\otimes \phi_2,\psi_1\otimes \psi_2)=(\phi_1,\psi_1)_1(\phi_2,\beta\psi_2)_2\label{tensprodpk}
\end{equation}
where $\beta=1$ if $(.,.)_1$ is even, $\chi_2$ if $(.,.)_1$ is odd and $(.,.)_2$ is even, $i\chi_2$ if $(.,.)_1$ and $(.,.)_2$ are both odd.
\item $\chi=\chi_1\otimes \chi_2$,
\item $J=J_1\chi_1^{|J_2|}\hat\otimes J_2\chi_2^{|J_1|}=J_1\otimes J_2\chi_2^{|J_1|}$,
\item $\Omega^1=\Omega^1_1\hat\otimes \pi(\A_2)+\pi(\A_1)\hat\otimes \Omega^1_2$.
\end{itemize}
\end{definition}
We stress that in the only case we will be considering, one of the two Krein spaces will be finite-dimensional, so we do not need to be more specific about topological tensor products.

The NCSM in Lorentzian signature is defined thanks to an algebraic background ${\cal B}$ which is the tensor product of one coming from a manifold, which we call ${\cal B}_M$, and a finite one ${\cal B}_F$. We will now describe these two backgrounds.

First, the manifold $M$ is a four-dimensional open anti-Lorentzian manifold (West-Coast convention). We suppose that there exists a global tetrad, i.e. a  pseudo-orthonormal frame $e=(e_0,\ldots,e_3)$, which is the condition  for $M$ to    admit a spin structure   \cite{geroch1}. Such a spin structure is defined by the trivial spinor bundle ${\cal S}=M\times S$, with $S=\CC^4$, and the choice of the following   gamma matrices:
\bea
\gamma_0=\pmatrix{0&1_2\cr 1_2&0};\gamma_k=\pmatrix{0&-\sigma_k\cr \sigma_k&0}, k=1,2,3\cr
\mbox{with }\sigma_1=\pmatrix{0&1\cr 1&0}, \sigma_2=\pmatrix{0&-i\cr i&0}, \sigma_3=\pmatrix{1&0\cr 0&-1}
\eea
This choice of $\gamma$ matrices permits to identify $\End(S)$ with the complex Clifford algebra $\CC l(1,3)$, and the map  $\rho: e_\mu\mapsto\gamma_\mu$ gives an irreducible representation of $\CC l(T_xM)$ on $S$. The spinor space $S$ carries a natural Krein product defined by
\begin{equation}
(\psi,\phi)_S=\psi^\dagger\gamma_0\phi
\end{equation}
The gamma matrices are all self-adjoint with respect to this product, and this property characterizes it up to a non-zero real factor \cite{part1}.  For any vector $v\in T_xM$, the hermitian form $(.,\rho(v)\cdot .)$ is positive  definite iff  $v$ lies in one half of the timelike cone, which we define to be the future cone. This defines a time orientation on $M$.

We still need to define the ``local'' chirality and real structure $\chi_S$ and $J_S$. The chirality operator $\chi_S$ is none other than the matrix $\gamma^5=i\gamma^0\ldots\gamma^3$ which in the chosen (chiral) representation is $\diag[I_2,-I_2]$. It satisfies $\chi_S^\times=-\chi_S$. The real structure $J_S$ is $\psi\mapsto\gamma_2\bar \psi$, where $\bar\psi$ is the complex conjugate of $\psi$ in the chosen basis. One can easily check that $J_S$ anticommutes with gamma matrices and satisfies $J_S^2=1$, $J_S^\times=-1$.  The collection of objects $({\cal S}, \rho, \chi_S, H_S, J_S)$ is the algebraic way to define a spin structure on $M$ (see \cite{algback} for details). 

The neutral component of the spin group ${\rm Spin}(1,3)^0$ contains by definition the elements $g$ such that:
\begin{enumerate}
\item $gg^\times=1$,
\item  $J_Sg=gJ_S$,
\item  $\chi_S g=g\chi_S$,
\item $g \Omega^1_S g^{-1}=\Omega^1_S$, where $\Omega^1_S$ is the vector space generated by the gamma matrices.
\end{enumerate}
Now it will turn out to be useful to notice that the last condition is automatically satisfied given the first three  in dimension $4$. 

The background ${\cal B}_M$ has algebra $\A_M=\tilde{\cal C}^\infty(M)_c$, which is generated by the constants and the smooth compactly supported real functions. The pre-Krein space ${\cal K}_M$ consists of smooth   spinor fields with compact support equipped with the indefinite product:
\be
(\phi,\psi)=\int_M(\psi(x),\phi(x))_S dx^0\ldots dx^3
\ee

The  chirality $\chi_M$ and charge conjugation $J_M$ are constant and defined by $(\chi_M \Psi)(x)=\chi_S\Psi(x)$, $(J_M\Psi)(x)=J_S\Psi(x)$. We won't use the bimodules of 1-forms in this paper. For details see \cite{algback}.

Let us now describe the finite background. The finite space ${\cal K}_F$ is identified with ${\cal K}_0\otimes I$ where ${\cal K}_0$ is the vector space generated over $\CC$ by all the different fermion species and $I$ is the vector space generated by the four symbols $R,L,{\bar R},{\bar L}$.  The space ${\cal K}_0$ is equipped with a preferred basis given by the symbols of the elementary fermion species (including generations and color: this is a $24$-dimensional space). This defines a canonical scalar product for which this basis is orthonormal. Similarly $I$ has preferred basis $(R,\ldots,\bar L)$ and is equipped with the associated scalar product. These canonical scalar products on ${\cal K}_0$ and $I$ are both written $\bra .,.\ket$,with no risk of confusion. The space ${\cal K}_F$ is also equipped with a Krein product with fundamental symmetry $\eta_F$. The finite algebra ${\cal A}_F$ is $\CC\oplus \HH\oplus M_3(\CC)$, though this specific  form will play no role here. More important is its representation: it is block-diagonal, that is:
\be
\pi_F(a)(\phi\otimes\sigma)=\pi_\sigma(a)\phi\otimes \sigma,\ \forall \phi\in {\cal K}_0, \sigma=R,L,\bar R,\bar L
\ee
The precise definition of $\pi_F$ can be found, with the same notation as we use, in \cite{algback} or \cite{BF1} for instance. It will play no role in this paper. We assume that:
\begin{enumerate}
\item \label{h1} $\eta_F=1\otimes \eta_I$ with $\eta_I=\diag[\eta_R,\eta_L,\eta_{\bar R},\eta_{\bar L}]$ a fundamental symmetry on $I$, 
\item \label{h2}$J_F(\phi\otimes \sigma)= \bar\phi\otimes\bar\sigma$, for all $\phi\in{\cal K}_0$ and $\sigma=R,L,\bar R,\bar L$, with the convention that  $\bar{\bar \sigma}=\epsilon_F\sigma$. 
\end{enumerate}
\begin{rem} The first hypothesis ensures that $\pi_F$ is $*$-representation. The second one is necessary in order to recover the correct representation of the gauge group.
\end{rem}

From \ref{h1} and \ref{h2} we easily obtain $\epsilon_F''=\kappa_F''=-1$.


It will be useful to have $J_F$ written in matrix form in the basis $(R,\bar L,\bar R, L)$. It is:
\be
J_F=1\otimes \pmatrix{0&0&\epsilon_F&0\cr
0&0&0&1\cr
1&0&0&0\cr
0&\epsilon_F&0&1
}\circ c.c.
\ee
Using this, we immediately see that in order to have $J_F\eta_F=\epsilon_F\kappa_F \eta_FJ_F$ as required by \eqref{compsigns}, we must have $\eta_{\bar R}=\epsilon_F\kappa_F\eta_R$ and $\eta_{\bar L}=\epsilon_F\kappa_F\eta_L$. Thus, to sum up the situation, hypotheses \ref{h1} and \ref{h2} imply that
\begin{equation}
\epsilon_F''=\kappa_F''=-1,\ \eta_{\bar R}=\epsilon_F\kappa_F\eta_R,\ \eta_{\bar L}=\epsilon_F\kappa_F\eta_L\label{releta}
\end{equation}
The Krein product on ${\cal K}_F$ is recovered through:
\be
(\phi\otimes\sigma,\phi'\otimes\sigma')_F=\bra\phi,\phi'\ket\bra\sigma,\eta_I\sigma'\ket
\ee
Here again the bimodule of 1-forms will play no role, so we ignore it. Let us now look at  ${\cal B}={\cal B}_M\hat\otimes {\cal B}_F$, the total background of the Standard Model. Its algebra is $\A=\tilde{\cal C}^\infty_c(M,\A_F)$ and its pre-Krein space is ${\cal K}={\cal C}_c^\infty(M,S\times {\cal K}_F)$. Let us look at the Krein product. It is obtained by integrating a product on $S\otimes {\cal K}_F$, which is, according to \eqref{tensprodpk}:
\bea
(\psi\otimes\phi\otimes\sigma,\psi'\otimes\phi'\otimes\sigma')&=&(\psi,\psi')_{S}\bra\phi,\phi'\ket\bra\sigma,\omega \sigma'\ket
\eea
where $\omega$ is an effective internal metric given by $\omega=\chi_I\eta_I=\diag[\eta_R,-\eta_{ L},-\eta_{\bar R},\eta_{\bar L}]$. The total chirality is 
\be
\chi=\chi_M\otimes \chi_F=\chi_M\otimes \id\otimes \chi_I
\ee
and the real structure is $J=J_M\chi_M^{|J_F|}\hat\otimes J_F\chi_F^{|J_M|}$.  In four spacetime dimensions and West-Coast convention, we have $\epsilon_M=\kappa_M''=1$, $\epsilon_M''=\kappa_M=-1$. Thus $J=J_M\chi_M\hat\otimes J_F\chi_F$, which can also  be written $J=J_M\otimes J_F\chi_F$. Hence we have
\bea
J(\Psi\otimes\phi\otimes R)&=&J_M\Psi\otimes\bar\phi\otimes \bar R\cr
J(\Psi\otimes\phi\otimes \bar R)&=&-\epsilon_FJ_M\Psi\otimes\bar\phi\otimes R\cr
J(\Psi\otimes\phi\otimes L)&=&-J_M\Psi\otimes\bar\phi\otimes \bar L\cr
J(\Psi\otimes\phi\otimes \bar L)&=&\epsilon_FJ_M\Psi\otimes\bar\phi\otimes L   \label{signJ}
\eea

\section{Statement and proof of the main result}
\subsection{Invariance under ${\cal U}_{J,\chi}$}
The fermionic action for the NCSM is generally given by
\be
S(A,\Psi)=(\Psi,D_A\Psi)\label{fa}
\ee
where $\Psi$ is an element of $\K$ and $D_A=D+A+JAJ^{-1}$ is a fluctuated Dirac operator. Note that $D_A$ is itself a Dirac operator, with the same commutation relations with $J$ and $\chi$ as $D$, and it can be argued that the true variable of the action is a compatible Dirac operator $D$, belonging to some subspace of $\D_\B$ which is invariant by $\Aut(\B)$ \cite{algback}. Note also that in the Euclidean setting the fermionic action can be written $S(D,\Psi)=(\Psi,JD\Psi)$ \cite{SMmix}. More generally, it would make sense to suppose the fermionic action to be of the form 
\be
S(D,\Psi)=(\Psi,QD\Psi)\label{genfermaction}
\ee
where $Q$ is a polynomial in $\chi$ and $J$. Indeed, $\chi$, $J$, as well as the Krein product $(.,.)$ are the only background objects at our disposal, thus \eqref{genfermaction} is the most general function of the variables $\Psi$ and $D$ that we can write down  which is at most of degree 1 in $D$ and quadratic in $\Psi$. Now it is immediate to observe that any action of this form is invariant under the transformation $\Psi\mapsto U\Psi$, $D\mapsto UDU^{-1}$, where $U$ belongs to the group 
$${\cal U}_{J,\chi}:=\{U\in\End(\K)|UU^\times=1, U\chi=\chi U, UJ=JU\}$$

Hence we see that the natural invariance group of the fermionic action, is larger than the automorphism group of $\B$. In particular it does not depend on the algebra and bimodule of 1-forms, and will stay the same for models beyond the Standard one which have the same fermion space, such as the $B-L$-extended SM \cite{algback} or Pati-Salam  \cite{ccvs}.

%
%
Now we want to ``solve the fermion  doubling problem'', that is, to find for each $x\in M$ a real \emph{physical subspace} $H_x$ of $V:=S\otimes {\cal K}_F$ which has the correct  dimension, i.e. $\dim_\RR(V)/4$, and to restrict the theory to the space ${\cal H}$ of \emph{physical fields} $x\mapsto \Psi(x)\in H_x$. Of course the space of physical fields has to be invariant under $\Aut(\B)$, but in view of the larger group of invariance of the fermionic action, it is very natural to postulate that ${\cal H}$ be invariant under ${\cal U}_{J,\chi}$ as well.

Let us look at some particular elements of this group.

\begin{lemma}
The following operators $U: \Psi\mapsto U\Psi$, $\Psi\in{\cal K}={\cal C}_c^\infty(M,S\times {\cal K}_F)$ belong to ${\cal U}_{J,\chi}$ when $(U\Psi)(x)$ is defined by:
\begin{enumerate}
\item $U\Psi(x)=\sqrt{\frac{{\rm vol}_{\theta^*g}(x)}{{\rm vol}_g(x)}}\Psi(\theta^{-1}(x))$, where $\theta$ is a diffeomorphism of $M$
\item $U\Psi(x)=(u\otimes \id)\Psi(x)$, where $u\in{\rm Spin}(1,3)^0$,
\item $U\Psi(x)=(\id\otimes u)\Psi(x)$, where $u\in {\cal U}_{J_F,\chi_F}$.
\end{enumerate}
\end{lemma}
For the proof, see \cite{algback}. Using invariance under the elements of the first kind, we see that $H_x$ does not depend on $x$.  Hence we want to find a real subspace $H$ of $V=S\otimes {\cal K}_F$, of dimension ${1\over 4}\dim_\RR V$, which is invariant under the group $G$ of Krein unitary operators on $V$, commuting with $J_x=J_S\otimes J_F\chi_F$ and $\chi_x=\chi_S\otimes \chi_F$. In the following sections we suppress the $x$ altogether, and we write $J,\chi$ instead of $J_x,\chi_x$. With these notations, we are going to prove the following result.

\begin{theorem}\label{th1}  There exists a $G$-invariant real subspace $H$ of $V$ of dimension $\dim_\RR(V)/4$ iff $\epsilon_F=-1$. In that case, there are two families of solutions, each parametrized by $S^1$. They are given by $H=(1+e^{i\varphi}J)({1\pm\chi})(V)$, $\varphi\in\RR$.
\end{theorem}

Note that the $S^1$-degree of freedom exactly corresponds to the ambiguity in the choice of $J$.
\subsection{Some preliminaries on Weyl spinors}
Recall that, in the chiral basis, the group $\Spin(1,3)^0$ is represented on $S$ by the matrices $\pmatrix{A&0\cr 0&\sigma_2\bar A\sigma_2}$ with $A\in SL_2(\CC)$. The spaces of Weyl spinors $S^+$ and $S^-$ contain respectively the vectors $\pmatrix{\phi\cr 0}$ and $\pmatrix{0\cr \phi}$. They are irreducible $\Spin(1,3)^0$-modules. Over $\CC$ they are the non-isomorphic $(1/2,0)$ and $(0,1/2)$-modules respectively. However, they are isomorphic over $\RR$.

\begin{lemma}\label{lem3} The group of real linear automorphisms of the $\Spin(1,3)^0$-modules $S^\pm$ is $\CC^*$.
\end{lemma}
\begin{demo} Let the  map $K : \CC^2\rightarrow \CC^2$ be $\RR$ and $SL_2(\CC)$-linear. Using the Lie algebra, $K$   commutes with the traceless matrices $\pmatrix{0&i\cr i&0}$ and $\pmatrix{0&1\cr 1&0}$, hence with their product, which is $i1_2$. Thus $K$ is $\CC$-linear, and the result follows from Schur's lemma.
\end{demo}

\begin{lemma}\label{lem4}
The set of $\RR$-linear $\Spin(1,3)^0$-isomorphisms from $S^+$ to $S^-$ is $\CC^*J_S$. 
\end{lemma}
\begin{demo}
The real structure $J_S$ is clearly an isomorphism over $\RR$ between $S^+$ and $S^-$, since  $\Spin(1,3)^0$ commutes with $J_S$ by its very definition. The result then follows using the previous lemma.
\end{demo}

In the sequel we write $(e_1^+,e_2^+,e_1^-,e_2^-)$  for the canonical  (chiral) basis of $S=S^+\oplus S^-$. 


\subsection{Proof of the theorem}
Writing down the general form of the elements of $G$ would involve $8\times 8$ block matrices, and would not turn out to be particularly useful. Instead we will work with special elements. First, $\Spin(1,3)^0\otimes 1$ and $1\otimes {\cal U}_{J_F,\chi_F}$ are obviously subgroups of $G$. The general form of  elements commuting with $J_F$ and $\chi_F$, with the basis order $(R,\bar L,\bar R,L)$ in $I$ is:
\be
u=\pmatrix{A&D&0&0\cr C&B&0&0\cr 0&0&\bar A&\epsilon_F\bar D\cr 0&0&\epsilon_F\bar C&\bar B}\label{genform}
\ee
with $A,B,C,D\in \End({\cal K}_0)$. Since $\eta_F=1\otimes [\eta_R,\eta_{\bar L},\eta_{\bar R},\eta_L]$ in this basis order, $u$ is Krein unitary iff the following conditions are met:
\bea
A^\dagger A+\eta_R\eta_{\bar L}C^\dagger C&=&1\cr
A^\dagger D+\eta_R\eta_{\bar L}C^\dagger B&=&0\cr
\eta_R\eta_{\bar L}D^\dagger D+B^\dagger B&=&1\label{eqUF}
\eea
Note that the 6 conditions boils down to only 3 using $\eta_R\eta_{\bar L}=\eta_{\bar R}\eta_{L}$.

In the sequel we will consider the subgroup  $G'$ of $G$  generated by 
\begin{enumerate}
\item  $\Spin(1,3)^0\otimes 1$, and  
\item\label{type2} the group $G''$ of endomorphisms $V$ of the form $1\otimes u$, where $u\in \End({\cal K}_F)$ is $[A,B,\bar A,\bar B]$ with $A,B$ unitary matrices of ${\cal K}_0$, which corresponds to the particular solution $C=D=0$ of \eqref{eqUF}.
\end{enumerate}

Let us define the 8 subspaces $H_\sigma^\pm:=S^\pm\otimes {\cal K}_0\otimes \sigma$, $\sigma=R,L,\bar R,\bar L$. Note that we consider them as real vector spaces. We clearly have:
\be
V=\bigoplus_{\pm,\sigma}H_\sigma^\pm
\ee
We call $p_\sigma^\pm$ the projections relative to this decomposition, which are $G'$-linear.
\begin{lemma} The real spaces $H_\sigma^\pm$  are irreducible $G'$-modules.
\end{lemma}
\begin{demo} It is clear that they are stable under $\Spin(1,3)^0$, and for $1\otimes u$ of the form \ref{type2}, we have, for all $\psi\in S^\pm$ and $\phi\in{\cal K}_0$:
\bea
(1\otimes u) \psi\otimes \phi\otimes R&=&\psi\otimes A\phi\otimes R\cr
(1\otimes u) \psi\otimes \phi\otimes L&=&\psi\otimes \bar B\phi\otimes L\cr
(1\otimes u) \psi\otimes \phi\otimes \bar R&=&\psi\otimes \bar A\phi\otimes \bar R\cr
(1\otimes u) \psi\otimes \phi\otimes \bar L&=&\psi\otimes B\phi\otimes \bar L\nonumber
\eea
which shows that $H_\sigma^\pm$ is stable under $1\otimes u$. Now let $W$ be a non-zero $G'$-submodule of $H_\sigma^\pm$, and let $w\not=0$ be an element of $W$. In the basis $(e_1^\pm,e_2^\pm)$ of $S^\pm$, the spin group elements are $SL_2(\CC)$ matrices. We can write $w$ in the form:
\be
w=e_1^\pm\otimes\phi_1\otimes\sigma+e_2^\pm\otimes \phi_2\otimes \sigma
\ee
In the $+$ case, acting with the spin group elements $A=\pmatrix{-1&0\cr 1&-1}$ and $A'=\pmatrix{-1&1\cr 0&-1}$, we see that 
\bea
w+Ae_1^+\otimes\phi_1\otimes\sigma+Ae_2^+\otimes\phi_2\otimes\sigma&=&e_2^+\otimes\phi_1\otimes\sigma\in W\cr
w+A'e_1^+\otimes\phi_1\otimes\sigma+A'e_2^+\otimes\phi_2\otimes\sigma&=&e_1^+\otimes\phi_2\otimes\sigma\in W\nonumber
\eea
In the $-$ case we replace $A$ and $A'$ by $\sigma_2\bar A\sigma_2$ and $\sigma_2\bar A'\sigma_2$ respectively, and we obtain the same result. Since $\phi_1$ and $\phi_2$ cannot both vanish, we see that there is a pure tensor in $W$. Now since $S^\pm$ is an irreducible \emph{real} $\Spin(1,3)^0-$module, we know that the real linear span of $Ae_1^\pm$ when $A$ varies in the spin group is $S^\pm$, and similarly with $Ae_2^\pm$. Thus we see that $W$ contains all vectors of the form $\psi\otimes\phi\otimes\sigma$, where $\phi=\phi_1$ or $\phi=\phi_2$ and $\psi$ is arbitrary in $S^\pm$. Now ${\cal K}_0$ is an irreducible real module for the unitary group. Consequently, if we act with $1\otimes U$ as above, we obtain that $\psi\otimes\phi\otimes\sigma\in W$ where $\psi$ and $\phi$ are both arbitrary, and this shows that $W=H_\sigma^\pm$.
\end{demo}
 
\begin{lemma}\label{lem5} For every $\sigma$, the $G'$-modules   $H_\sigma^+$ and $H_{\bar\sigma}^-$ are isomorphic. The isomorphisms from $H_\sigma^+$ to $H_{\bar\sigma}^-$ are of the form $\lambda J_S\otimes J_F$, $\lambda\in\CC^*$. No other distinct modules among the $H_\sigma^\pm$ are isomorphic. 
\end{lemma}
\begin{demo}
The given maps are quickly seen to be $\RR$-linear isomorphisms. Let us prove that they are $G'$ linear. For this, consider $A\in \Spin(1,3)^0$ and $u\in{\cal U}_{J_F,\chi_F}$. Then $J_S\otimes J_F$ commutes with $A\otimes 1$ since $J_S$ commutes with $A$, and it also commutes with $1\otimes u$ since $J_F$ commutes with $u$.
%
%
%
The uniqueness up to a scalar of the above isomorphisms is immediate from the fact that $\Aut(H_\sigma^+)=\CC^*$, which we now need to prove. This follows from Schur's lemma once we have proved that these automorphisms are all $\CC$-linear. This is proven with the exact same method as in lemma \ref{lem3}. Finally, we must prove that no other isomorphism exists between the $H_\sigma^\pm$. It is obvious that $H_\sigma^\pm\not\simeq H^{\cdot}_{\sigma'}$ if $\sigma'\not=\sigma$ and $\sigma'\not=\bar\sigma$, since $1\otimes u\in G''$ will act by $A$ or $\bar A$ in one case and by $B$ or $\bar B$ in the other.  Now let us suppose $\theta : H_\sigma^+\rightarrow H_\sigma^-$ is a $G'$-linear map. Let us decompose $H_\sigma^+$ and $H_\sigma^-$ into irreducible $\Spin(1,3)^0$-modules: $H_\sigma^+=\bigoplus_f S^+\otimes f\otimes \sigma$, and $H_\sigma^-=\bigoplus_f S^-\otimes f\otimes \sigma$, with $f$ running over a basis of elementary fermions. Since $\theta$ is $\Spin(1,3)^0$-linear, the ``matrix elements'' $\theta_{ff'}$ determined by this decomposition are all $\CC$-antilinear  by lemma \ref{lem4}. Thus $\theta$ is $\CC$-antilinear. But if we now decompose $H_\sigma^+$ and $H_\sigma^-$ into a sum of irreducible $G''$-modules, we likewise find that $\theta$ is $\CC$-linear. Hence $\theta=0$. Using the same method we find that $H_\sigma^\pm$ and $H^\pm_{\bar\sigma}$ are not isomorphic.
\end{demo}

Gathering the isomorphic summands, we obtain the decomposition of $V$ into isotypical $G'$-components.
\be
V=(H^+_R\oplus H^-_{\bar R})\oplus (H^+_{\bar L}\oplus H^-_L)\oplus  (H^+_L\oplus H^-_{\bar L})\oplus (H^+_{\bar R}\oplus H^-_R)
\ee
Let $W$ be an irreducible $G'$-submodule of $V$. The projections $p_\sigma^\pm$ restricted to $W$ are isomorphisms or vanish. Since they cannot all vanish, we see that $W$ is isomorphic to $H^+_R,H^-_{L}, H^+_L$ or $H^-_{ R}$, the four cases corresponding to the isotypical components, and we call them respectively the right-even, left-odd, left-even and right-odd cases.
\begin{propo}\label{irredmod} Let $W$ be an irreducible $G'$-submodule of $V$.
\begin{enumerate} 
\item\label{claim1}   $W$ is right-even iff   there is a pair $\alpha=(\alpha_+^R,\alpha_-^{\bar R}) \in\CC^2\setminus\{(0,0)\}$  such that $W=W^{+,R}_\alpha:=\{\sum\alpha_+^R\psi_R\otimes\phi\otimes R+\alpha_-^{\bar R}J_S\psi_R\otimes \bar\phi\otimes \bar R|\psi\in S^+,\phi\in {\cal K}_0\}$,
\item $W$ is left-odd iff there is a pair $\alpha=(\alpha_-^L,\alpha_+^{\bar L}) \in\CC^2\setminus\{(0,0)\}$  such that $W=W^{-,L}_\alpha:=\{\sum\alpha_-^L\psi_L\otimes\phi\otimes L+\alpha_+^{\bar L}J_S\psi_L\otimes \bar\phi\otimes \bar L|\psi_L\in S^-,\phi\in {\cal K}_0\}$,
\item $W$ is left-even iff   there is a pair $\alpha=(\alpha_+^L,\alpha_-^{\bar L}) \in\CC^2\setminus\{(0,0)\}$  such that $W=W^{+,L}_\alpha:=\{\sum\alpha_+^L\psi_R\otimes\phi\otimes L+\alpha_-^{\bar L}J_S\psi_R\otimes \bar\phi\otimes \bar L|\psi\in S^+,\phi\in {\cal K}_0\}$,
\item $W$ is right-odd iff there is a pair $\alpha=(\alpha_-^R,\alpha_+^{\bar R}) \in\CC^2\setminus\{(0,0)\}$  such that $W=W^{-,R}_\alpha:=\{\sum\alpha_-^R\psi_L\otimes\phi\otimes R+\alpha_+^{\bar R}J_S\psi_L\otimes \bar\phi\otimes \bar R|\psi_L\in S^-,\phi\in {\cal K}_0\}$\end{enumerate}
\end{propo}
\begin{demo}
 Let us suppose $W$ is right-even, the other cases being entirely similar. Then we know that all the projections $p_\sigma^\pm$ vanish on $W$, except at least one among $p_R^+$ and $p^-_{\bar R}$. Suppose $p_R^+$ does not vanish. It is then an isomorphism. Let $f : H^+_R\rightarrow W$ be its inverse. Then every $v\in W$ is of the form $f(\sum \psi\otimes \phi\otimes R)$, $\psi\in S^+,\phi\in{\cal K}_0$. We know by lemma \ref{lem5} that $p^-_{\bar R}\circ f$ is of the form $\lambda J_S\otimes J_F$, with $\lambda\in\CC$. Thus every $v\in W$ is of the form
\bea
v=f(\sum \psi\otimes \phi\otimes R)&=&(p^+_R+p^-_{\bar R})\circ f(\sum \psi\otimes \phi\otimes R)\cr
&=&\sum \psi\otimes\phi\otimes R+\lambda J_S\psi\otimes\bar\phi\otimes \bar R\nonumber
\eea
The pair  $(1, \lambda)$ is non-vanishing. If $p^+_R$ vanishes on $W$, then we do the same reasoning with $p^-_{\bar R}$. In the end we obtain a non-vanishing pair such that $W$ has the required form. Conversely every $W$ as in the statement is clearly a $G'$-module isomorphic to $H^+_R$.

The 3 other cases are similar, except that we sometimes have to absorb signs coming from \eqref{signJ} in the definition of the coefficients $\alpha_\pm^\sigma$.
\end{demo}
The pair $\alpha$ in this proposition is not unique. Let us define the following equivalence relation on $\CC^2\setminus\{(0,0)\}$:
\be
(\alpha_1',\alpha_2')\sim (\alpha_1,\alpha_2)\Leftrightarrow \exists\lambda\in\CC^*, \alpha_1'=\lambda\alpha_1,\alpha_2'=\bar\lambda\alpha_2
\ee
\begin{lemma}\label{lemsim}
Let $W_\alpha^{\pm,\sigma}$ and $W_{\alpha'}^{\pm,\sigma}$ be two irreducible $G'$-modules. Then $W_{\alpha'}^{\pm,\sigma}=W_\alpha^{\pm,\sigma}$ iff $\alpha'\sim\alpha$.
\end{lemma}
\begin{demo}
We work in the right-even case, the others being similar. Let $\alpha=(a,b)$, $\alpha'=(a',b')$. An element of $W_\alpha^{+,R}$ can be uniquely written as:
\bea
v&=&\sum_{i,f}\big(a\lambda_{i,f}e_i^+\otimes f\otimes R+b\bar\lambda_{i,f}J_Se_i^+\otimes f\otimes\bar R\big)\nonumber
\eea
We thus see that $a'e_i^+\otimes f\otimes R+b'J_Se_i^+\otimes f\otimes\bar R$ can be written in this way iff $\exists\lambda\in\CC$ such that $a'=\lambda a$ and $b'=\bar\lambda b$.
\end{demo}

The space $H$ we are looking for is a $G$-module, hence a $G'$-module, and as such it is the direct sum of irreducible sub-modules of the   kind described in proposition \ref{irredmod}. For dimensional reasons, we need two of them. Now  a particular solution of \eqref{eqUF} is $u=1_{{\cal K}_0}\otimes r$ where $r=\pmatrix{\cos t&-\sin t\cr \sin t&\cos t}$   if $\eta_R\eta_{\bar L}=1$ and $r=\pmatrix{\cosh t&\sinh t\cr \sinh t&\cosh t}$ if  $\eta_R\eta_{\bar L}=-1$. Hence the elements of $1\otimes {\cal U}_{J_F,\chi_F}$ can mix the letters $(R,\bar L)$ on one hand, and $(\bar R,L)$ on the other. Thus it is clear that $H$ must be the sum of a left-even/odd and a right-even/odd module. We will now show that the ``parity'' of the two modules must be opposite. For this, let us  introduce a particular class of elements of $G$.
\begin{lemma} The operator $\gamma_\mu\otimes T$ is in $G$ iff $\{\chi_F,T\}=0$, $[J_F,T]=0$ and $T^\times T=-(\gamma_\mu)^2$.
\end{lemma}
The proof is immediate. Thus $T$ has the block-matrix form
\be
T=\pmatrix{0&0&\epsilon_F\bar A&\bar B\cr 0&0&\bar C&\epsilon_F\bar D\cr A&B&0&0\cr C&D&0&0}
\ee
with 
%
\bea
 \eta_R\eta_{\bar R}A^\dagger A+\eta_R\eta_LC^\dagger C&=&-(\gamma_\mu)^2\cr
\eta_{\bar R}A^\dagger B+\eta_LC^\dagger D&=&0\cr
\eta_{\bar L}\eta_{\bar R}B^\dagger B+\eta_{\bar L}\eta_{L}D^\dagger D&=&-(\gamma_\mu)^2\label{eqT}
\eea
A particular solution to \eqref{eqT} is $A=D=0$, $C=B=1$ and:
\begin{itemize}
\item $\gamma_\mu=\gamma_0$ if $\eta_R\eta_{\bar R}=\epsilon_F\kappa_F=-1$,
\item $\gamma_\mu=\gamma_1$ if $\eta_R\eta_{\bar R}=\epsilon_F\kappa_F=1$.
\end{itemize}
We write this solution $T=\gamma_\mu\otimes 1\otimes \tau$, where $\tau$ exchanges the letters $R$ and $L$. Hence we have (discarding the ${\cal K}_0$-factor, which plays no role):
\bea
\gamma_\mu\otimes \tau(\alpha^R_+\psi_R\otimes R+\alpha_-^{\bar R}J_S\psi_R\otimes \bar R)&=&\alpha^R_+\gamma_\mu\psi_R\otimes L-\alpha_-^{\bar R}J_S\gamma_\mu\psi_R\otimes \bar L\cr
&&\label{eqbiz}
\eea
Thus we see that if we take $H$ to be the sum of a right-even and left  module, then the left module must  be odd, since $\gamma_\mu\psi_R\in S^-$. Similarly a right-odd module must be associated with a left-even one. Hence we find two possible solutions so far:
\bea
H&=&W_\alpha^{+,R}\oplus W_\beta^{-,L},\mbox{ or }\cr
H&=&W_\alpha^{-,R}\oplus W_\beta^{+,L}\nonumber
\eea
Suppose we are in the first case.  Equation \eqref{eqbiz} yields conditions on the pairs of complex numbers $\alpha=(\alpha_+^R,\alpha_-^{\bar R})$ and $\beta=(\alpha_-^L,\alpha_+^{\bar L})$: it shows that $(\alpha_-^L,\alpha_+^{\bar L})$ and $(\alpha^R_+,-\alpha_-^{\bar R})$ define the same left-odd module. We can thus take them to be equal without loss of generality. Thus we have
\be
H=W_{(a,b)}^{+,R}\oplus W_{(a,-b)}^{-,L}
\ee
%

We will now use another particular solution to \eqref{eqT}. If $\epsilon_F\kappa_F=1$ we can take $B=C=0$, $A=D=1$, with $\mu=1,2$ or $3$. If $\epsilon_F\kappa_F=-1$ we have the same solution but with $\mu=0$. Hence we have the operator $\gamma_\mu\otimes 1\otimes \tau$ where this time $\tau(R)=\bar R$, $\tau(\bar R)=\epsilon_F R$ and similarly with $L$. We obtain $(\gamma_\mu\otimes T)(W_{(a,b)}^{+,R})=W_{(-\epsilon_Fb,a)}^{+,R}$, hence we must have $W_{(a,b)}^{+,R}=W_{(-\epsilon_Fb,a)}^{+,R}$. Thus $(a,b)\sim(-\epsilon_Fb,a)$ by lemma \ref{lemsim}. This means that there is a $\lambda\in\CC^*$ with $a=-\lambda \epsilon_F b$ and $b=\bar\lambda a$. We see that $(1+\epsilon_F|\lambda|^2)b=0$. Since $a=b=0$ is impossible, we obtain that $\epsilon_F=-1$ and $|\lambda|=1$. Hence $a$ and $b$ have the same modulus $r$, and since $(re^{i\theta},re^{i\phi})\sim (1,e^{i(\phi-\theta)})$, we see that:
\be
H=W_{(1,e^{i\varphi})}^{+,R}\oplus W_{(1,-e^{i\varphi})}^{-,L}
\ee
The elements of $H$ are then of the form
\bea
v&=&\sum_{\psi_R\in S^+,\phi\in {\cal K}_0}\psi_R\otimes \phi\otimes R+e^{i\varphi}J_S\psi_R\otimes\bar\phi\otimes\bar R\cr
&&+\sum_{\psi_L\in S^-,\phi'\in {\cal K}_0}\psi_L\otimes \phi'\otimes L-e^{i\varphi}J_S\psi_L\otimes\bar\phi'\otimes\bar L\cr
&=&\sum_{\psi_R\in S^+,\phi\in {\cal K}_0}\psi_R\otimes \phi\otimes R+e^{i\varphi}J(\psi_R\otimes\phi\otimes R)\cr
&&+\sum_{\psi_L\in S^-,\phi'\in {\cal K}_0}\psi_L\otimes \phi'\otimes L+e^{i\varphi}J(\psi_L\otimes\phi'\otimes L)\label{formH}
\eea
Thus $H=(1+e^{i\varphi}J)({1-\chi\over 2})(V)$, and this immediately shows that $H$ is stable by the whole group $G$. The odd case is entirely similar and we find $H={(1+e^{i\varphi}J}({1+\chi\over 2})(V)$. This proves the theorem.

\section{Generalized Majorana-Weyl conditions and KO-dimension}

Remember we had $\epsilon_F''=\kappa_F''=-1$. Consulting table \ref{tab1} we see that $\epsilon_F=-1$ iff the KO-dimension $n_F=2$. Since the KO-dimension of the manifold $n_M=6$ we can restate theorem \ref{th1} by saying that there exists a $G$-invariant subspace of dimension $\dim_\RR(V)/4$ iff the KO-dimension of the total algebraic background is $0$. This also amounts to say that $\epsilon=\epsilon''=1$.

Moreover,  introducing an inocuous factor of $1/4$, the space $H$ is $p_\pm(V)$ where $p_\pm={1+e^{i\varphi}J\over 2}{1\pm\chi\over 2}$. Using $\epsilon=\epsilon''=1$ we see that $p_\pm$ is the product of two commuting projectors (the converse is also true). Of course ${1\pm\chi\over 2}$ projects on the $\pm 1$-eigenspaces of $\chi$, and ${1\pm e^{i\varphi}J\over 2}$ projects on the $\pm e^{-i\varphi}$-``eigenspaces''\footnote{These are not really eigenspaces of the $\RR$-linear operator $J$, which can only have a $\pm 1$-eigenspaces, these are the space where $J$ and the real operator of multiplication by $e^{-i\varphi}$ coincide.} of $J$. Thus, the solution of the fermion quadrupling problem, assuming ${\cal U}_{J,\chi}$-invariance, is given by the generalized ``Majorona-Weyl conditions"
\bea
\chi\psi&=&\pm \psi\cr
J\psi&=&e^{-i\varphi}\psi
\eea
This furnishes a  converse to Barrett's solution with the hypothesis of invariance under ${\cal U}_{J,\chi}$. Note that without this hypothesis, other solutions exist. Indeed, let $\alpha,\beta,\gamma,\delta$ be four arbitrary complex numbers. Then the real space $H_{\alpha,\beta,\gamma,\delta}$ comprising vectors of the form\footnote{We thank Nadir Bizi for this observation.}
\bea
\alpha \psi_R\otimes \phi\otimes R+\beta J_M\psi_R\otimes \bar \phi\otimes \bar R+\gamma \psi_L\otimes \phi'\otimes L+\delta J_M\psi_L\otimes\phi'\otimes \bar L
\eea
is quickly seen to be invariant under the gauge group of the Standard Model, or for that matter, any model with the same finite space and a finite algebra which does not mix letters $R,L,\bar R,\bar L$. This space has the correct dimension and could also be taken to be a solution of the fermion quadrupling problem.

Let us close this section by observing that if we insert the Majorana-Weyl condition into the usual form \eqref{fa} of the fermionic action, we find conditions on $\kappa$ and $\kappa''$ for this action not to vanish. These conditions vary according to whether the fermionic variables are taken to be commuting or anti-commuting. In order to treat the two cases on the same footing, let us set $s=1$ if the fermion variables commute, and $s=-1$ if they anti-commute. We then find that:
\be
(\psi,D\psi)\not\equiv 0\Rightarrow \kappa''=-1\mbox{ and }\kappa=s\label{kappacond}
\ee
This is equivalent to say that $m=6$ if $s=1$ and $m=2$ if $s=-1$.  Since the metric dimension  of the manifold is $m_M=4$, we find that $m_F=2$ if $s=1$ and $m_F=6$ if $s=-1$. It is interesting to note that this last result is in agreement (modulo 8) with the number of compact dimensions in String Theory.

\section{Projection on the physical subspace and the fermionic action}
One can  see easily that ${1\pm e^{i\varphi}J\over 2}{1\pm \chi\over 2}$ are 4 projectors which sum to 1 and such that the product of any two of them vanish. Let us introduce the following notation:
\be
p^a_b={1\over 4}({1+ae^{i\varphi}J})(1+b\chi)
\ee
where $a,b=\pm 1$.

The traditional solution of the fermion quadrupling problem is to restrict the field $\psi$ to have values in the chosen physical subspace, say $p_1^1V$. However, it would be equivalent to suppose that $\psi$ has value in $V$, but that the fermionic action depends on $\psi$ only through $p_1^1\psi$. In order to formulate this idea more precisely, and to treat commuting and anti-commuting fermionic variables simultaneously, let us introduce the real bilinear form
\be
(\phi,\psi)_r:={1\over 2}((\phi,\psi)+(\psi,\phi))
\ee
If $s=1$ this is just the real part of $(.,.)$, however if $s=-1$ this interpretation cannot be maintained since the action has value in a Grassmann algebra. In any case, the usual form of the fermionic action is $S_f(D,\psi)=(\psi,D\psi)_r$, and this can be generalized to 
\be
S_f(D,\psi)=(\psi,QD\psi)_r\label{genfa}
\ee
where $Q$ is any polynomial in $\chi$ and $J$. If $B$ is a real linear operator we will write $B^+$ for the adjoint of $B$ with respect to $(.,.)_r$. Let us note that if $B$ is  $\CC$-linear then $B^+=B^\times$, whereas if $B$ is antilinear, $B^+=sB^\times$, since we have 
\bea
(\phi,B\psi)_r&=&{1\over 2}\big((\phi,B\psi)+(B\psi,\phi)\big)\cr
&=&{s\over 2}\big((\psi,B^\times \phi)+(B^\times \phi,\psi) \big)\cr
&=&s(B^\times \phi,\psi)_r
\eea
Now we see that if we take the action to be 
\bea
S_f(D,\psi)&=&(p_1^1\psi,Dp_1^1\psi)_r\label{physaction}
\eea
we obtain
\bea
16S_f(D,\psi)&=&((1+ae^{i\varphi}J)(1+b\chi)\psi,D(1+ae^{i\varphi}J)(1+b\chi)\psi)_r\cr
&=&(\psi,(1+b\kappa''\chi)(1+sa\kappa e^{i\varphi}J)(1+ae^{i\varphi}J)(1-b\chi)D\psi)_r\cr
&=&0, \mbox{ unless }\kappa''=-1, \kappa s=1
\eea
We thus find the same conditions as in \eqref{kappacond}. Moreover, if these conditions are met, then $J^\times=\kappa J=s J$, thus we have $J^+=J$, while $\chi^+=\epsilon''\kappa''\chi=-\chi$. We can then deduce  that
\be
(p^a_b)^+=p^{ a}_{-b}
\ee
Using $p^a_bD=Dp^a_{-b}$, we see that the action can also be written
\be
S_f(D,\psi)=(\psi,p_{-1}^1D\psi)_r\label{physaction2}
\ee
which has the form \eqref{genfa}. To see how particular this action is with respect to the general form,  we first note that thanks to the commutation relations among $J$ and $\chi$, as well as $J^2$ and $\chi^2=$ constants, the  algebra of polynomials $\CC[J,\chi]$ is four dimensional. Moreover, since 
\be
\left|\matrix{1&1&e^{i\varphi}&e^{i\varphi}\cr
1&1&-e^{i\varphi}&-e^{i\varphi}\cr
1&-1&e^{i\varphi}&-e^{i\varphi}\cr
1&-1&-e^{i\varphi}&e^{i\varphi}
}\right|\not=0
\ee
the projectors $p^a_b$ form a basis of $\CC[J,\chi]$. Hence $Q$ can  be written as
\be
Q=\sum_{a,b}\pi^b_a p_b^a
\ee
with $\pi^b_a\in\CC$, and we easily obtain
\bea
(\Psi,QD\psi)_r&=&2 \sum_{a,b}{\rm Re}(\pi_a^b)(p^a_{-b}\psi,Dp^a_{-b}\psi)_r
\eea
%
%
Thus the action \eqref{physaction}, up to a constant, is obtained when we set the real part of three out  of the four numbers $\pi_a^b$ to $0$. The constant can then be absorbed in a redefinition of $\psi$. This is a form of fine-tuning, but choosing the traditional  action $(\psi,D \psi)$  among all the possible action is also fine-tuning, and we would still have to restrict to the physical subspace. Hence we think that postulating \eqref{physaction} is a more economical solution. Note however that with this solution, the extra degrees of freedom are still there, even if they do not interact with gauge bosons. Let us conclude by remarking that if we take $e^{i\varphi}=1$ and  develop \eqref{physaction2}, we obtain a particularly symmetrical formula for the action, which is valid for both $s=1$ and $s=-1$: 

\begin{eqnarray}
S(D,\psi)=&\frac{1}{8}[(\psi,D\psi)+(D\psi,\psi)+(D\psi,J\psi)+(J\psi,D\psi)\cr
&+(\chi\psi,D\psi)+(D\psi,\chi\psi)+(D\psi,\chi J\psi)+(\chi J\psi,D\psi)]
\end{eqnarray}

Moreover, we can add to the action a term $(p'\psi,p'\psi)_r$, where $p'=1-p_{-1}^1$. This way $p'\psi$ is treated as an auxiliary field which is set to zero by the equations of motion.  

\section{Acknowledgements}
We thank Fedele Lizzi and Maxim Kurkov for their interesting and useful comments on a first version of this paper.
\bibliographystyle{unsrt}
\bibliography{../generalbib/SSTbiblio}

\end{document}